\documentclass[aps,prl,twocolumn]{revtex4}

\usepackage{graphics}
\usepackage{epsfig}

\usepackage{color}
\usepackage{amsmath}

\def\bra#1{\left\langle#1\right|}
\def\ket#1{\left|#1\right\rangle}

\begin{document}

\title{Generic Wavefunction Description of Fractional Quantum Anomalous Hall States and Fractional Topological Insulators}

\author{Xiao-Liang Qi$^{1}$}
\affiliation{
$^1$Department of Physics, Stanford University, Stanford, CA 94305, USA
}

\date{\today}

\begin{abstract}
%In this paper, we propose a systematical approach to construct topologically ordered states in two-dimensional lattice models. By making use of the concept of one-dimensional Wannier functions, both chiral topological states named as fractional quantum anomalous Hall states and non-chiral topological states named as fractional topological insulators can be constructed as generalizations of ordinary fractional quantum Hall states. Our results
%Fractional quantum Hall states are well-known topologically ordered states realized in two-dimensional electron gas with a strong perpendicular magnetic field. In this paper,
We propose a systematical approach to construct generic fractional quantum anomalous Hall (FQAH) states, which are generalizations of the fractional quantum Hall states to lattice models with zero net magnetic field and full lattice translation symmetry. %By making use of the concept of one-dimensional Wannier functions. %Such states are topologically equivalent to fractionally quantum Hall states but are defined for systems without orbital magnetic field, which are named as fractional quantum anomalous Hall states.
Local and translationally invariant Hamiltonians can also be constructed, for which the proposed states are unique ground states. Our result demonstrates that generic chiral topologically ordered states can be realized in lattice models, without requiring magnetic translation symmetry and Landau level structure. We further generalize our approach to the time-reversal invariant analog of fractional quantum Hall states--fractional topological insulators, and provide the first explicit wavefunction description of fractional topological insulators in the absence of spin conservation.
\end{abstract}

%\pacs{05.10.Cc,75.10.Jm,71.10.-w}

\maketitle

%Outline:
%
%1. Intro. Motivation and review. FQH on lattice models without magnetic field. Analytical description.
%
%2. Wannier function approach. $C_1=1$ case, Wannier function continuity. Map to Landau level case.
%
%3. Laughlin state and Pseudo-potential Hamiltonians. Coherent state basis and locality.
%
%4. Generalization to Jack polynomial.
%
%5. Discussion. Generalization to higher $C_1$ and FQSH.

{\it Introduction}--Topological states of matter are quantum states which are distinguished from ordinary states by topological properties, rather than more conventional properties such as symmetry. The first examples of topological states of matter discovered in nature are the integer and fractional quantum Hall states\cite{klitzing1980,tsui1982} which are realized in two-dimensional electron gas systems in a strong perpendicular magnetic field. The integer quantum Hall (IQH) state is characterized by a Chern number of the geometrical gauge field defined in the magnetic Brillouin zone\cite{thouless1982} or parameter space of twist boundary conditions\cite{niu1985,avron1985}, which determines the integer $n$ in the quantized Hall conductance $\sigma_{xy}=ne^2/h$. In a work in 1988\cite{haldane1988}, F. D. M. Haldane proposed the first realization of the IQH state in a band insulator model without net magnetic field. During recent years, several semiconductor systems have been proposed\cite{qi2005,liu2008B,yu2010} which may realize such an IQH state without Landau levels, named as the quantum anomalous Hall (QAH) state.

Based on the understanding of QAH states, a natural question is whether one can also find fractional quantum anomalous Hall (FQAH) states, which are fractional quantum Hall (FQH)  states without Landau levels. Different from IQH states, FQH states necessary require electron-electron interaction, which makes the generalization to the systems without Landau levels more difficult. In ordinary FQH states the kinetic energy of electrons are quenched due to the flat Landau level, so that the electron interaction effect can be significant and lead to topological nontrivial states. On the contrary, in a QAH system the energy dispersion is in general not flat, so that if we consider the interaction effect, there is competition between kinetic energy and interaction energy which usually disfavors the topological nontrivial states. Recently, specific QAH models with almost flat band dispersion and nontrivial Chern number has been constructed\cite{sun2010}. Numerical evidences of FQAH states have been found in such flat band models when interaction is considered\cite{wang2011,sheng2011}. However, the understanding of FQH states based on wavefunctions such as the Laughlin wavefunction\cite{laughlin1983} cannot directly apply to FQAH states, since the the single-particle and many-body wavefunctions in the QAH system are defined on lattice and cannot be written as analytic functions.

In this Letter, we propose a systematical way to describe the FQAH states by constructing model wavefunctions. We show that one-dimensional maximally localized Wannier functions can be defined in QAH states, which plays the same role as that of the Landau level wavefunctions in IQH and FQH states. Based on the Wannier function basis, we construct the analogue of Laughlin wavefunctions in FQAH, and also obtain the analogue of the pseudo-potential Hamiltonians\cite{haldane1983,trugman1985} of which these wavefunctions are exact ground states. Once such a one-to-one mapping between Wannier functions in QAH and Landau level wavefunctions in QH is defined, {\it each wavefunction constructed in FQH has a counterpart in the FQAH}. This demonstrates that the physics of FQH does not rely on any special property of the Landau level problem, such as the wavefunctions being analytic functions, and the magnetic translation symmetry. Instead, fractionalized topological states exist generically in a flat (or nearly flat) band with a nontrivial Chern number.

{\it QAH states and one-dimensional Wannier functions.--} The QAH state is a band insulator described by a Hamiltonian
\begin{align}
H=\sum_{\bf k}c_{\bf k}^\dagger h({\bf k})c_{\bf k}\label{HQAH}
\end{align}
The single particle Hamiltonian $h({\bf k})$ is a $N\times N$ Hermitian matrix for a system with $N$ bands. Denote the eigenstates of the single particle Hamiltonian $h({\bf k})$ as $\ket{n,{\bf k}},~n=1,2,...,N$ with eigenvalue $E_n$, the Hall conductance of the system is determined by the first Chern number\cite{thouless1982} $C_1=\int d^2{\bf k}f_{xy}({\bf k})$ with $f_{xy}({\bf k})=\partial_x a_y-\partial_y a_x$ and $a_i({\bf k})=-i\sum_{E_n<0}\bra{n,{\bf k}}\partial_i\ket{n,{\bf k}},~i=x,y$. The QAH state is described by a nontrivial Chern number. For a system with $C_1=0$, the wavefunction of states $\ket{n,{\bf k}}$ can be taken as single-valued in the Brillouin zone, from which one can construct the Wannier function basis by a Fourier transform $\ket{W_{n\bf x}}=\frac1{\sqrt{L_xL_y}}\sum_{{\bf k},m}e^{i{\bf k}\cdot {\bf x}}U_{nm}(k)\ket{m,{\bf k}}$ with $U_{nm}(k)$ some smooth unitary transformation. On the contrary, for QAH states with $C_1\neq 0$, it is well-known that Wannier function localized along both $x$ and $y$ directions cannot be defined since the wavefunction can not be taken as single-valued through the Brillouin zone.\cite{thouless1984,thonhauser2006} Instead, one-dimensional (1D) Wannier functions can be defined which are eigenstates of $k_y$ and are maximally localized along $x$ direction\cite{soluyanov2011,soluyanov2011b,yu2011}. For each fixed $k_y$, all states with momentum $k_y$ form a one-dimensional subsystem with the Hamiltonian $H_{\rm 1D}(k_y)=\sum_{k_x}c_{k_xk_y}^\dagger h(k_x,k_y)c_{k_xk_y}$. For one-dimensional systems there is no obstruction in getting localized Wannier function. Maximally localized Wannier functions can be obtained as eigenstates of the projected position operator ${X}=P_-xP_-$ with $x$ the x-direction coordinate operator and $P_-$ the projection to occupied bands.\cite{kivelson1982}

For simplicity, we consider a QAH system with only one occupied band denoted by $\ket{k_x,k_y}$. The Berry's phase gauge field $a_i=-i\bra{k_x,k_y}\partial_i\ket{k_x,k_y}$ with $i=x,y$. One can always make a gauge choice $a_y=0$, in which case the explicit form of the maximally localized Wannier function is
\begin{align}
\ket{W(k_y,x)}&=L_x^{-1/2}\sum_{k_x}e^{-i\int_0^{k_x} a_x(p_x,k_y)dp_x}\nonumber\\
&\cdot e^{-ik_x\left({x-\frac{\theta(k_y)}{2\pi}}\right)}\ket{k_x,k_y}\label{Wannier1d}
\end{align}
with $\theta(k_y)=\int_0^{2\pi}a_x(p_x,k_y)dp_x$.
$x\in{\rm Z}$ labeles the lattice sites. The phase factor with $e^{i\theta(k_y) k_x/2\pi}$ guarantees that the Bloch function is periodic for $k_x\rightarrow k_x+2\pi$. In the gauge transformation $\ket{k_x,k_y}\rightarrow e^{i\varphi(k_x,k_y)}\ket{k_x,k_y}$, the Wannier function is gauge invariant up to an overall phase: $\ket{W(k_y,x)}\rightarrow e^{i\varphi(0,k_y)}\ket{W(k_y,x)}$. It can be verified directly that the center of mass position of the Wannier function $\ket{W(k_y,x)}$ is given by \begin{align}
\left\langle \hat{x}\right\rangle=\bra{W(k_y,x)}\hat{x}\ket{W(k_y,x)}=x-\theta(k_y)/2\pi,\nonumber
\end{align} so that $\theta(k_y)/2\pi$ is the shift of the Wannier function away from the lattice site, {\it i. e.} the charge polarization.\cite{kingsmith1993}

\begin{figure}[htbp]
\centerline{\includegraphics[width=3.5in]{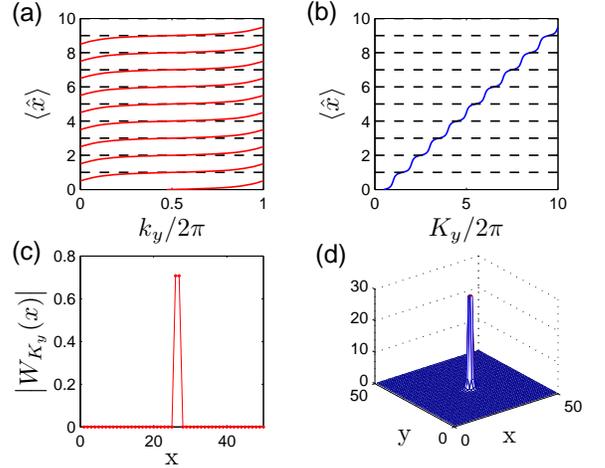}}
\caption{(Colour online) (a) The center-of-mass position $\left\langle\hat{x}\right\rangle$ of the Wannier functions versus $k_y$. (b) $\left\langle\hat{x}\right\rangle$ versus the extended wavevector $K_y$ defined in Eq. (\ref{Kydef}). (c)The profile of a Wannier function. (d) The profile of the coherent state wavefunction defined in Eq. (\ref{coherentstate}). All the results are calculated for the two-band model (\ref{HDirac}) with the parameters $M=-1,B=1/2$ on a $50\times 50$ lattice.}\label{fig1}
\end{figure}

Since the polarization $\theta(k_y)$ is determined by the Wilson loop of the Berry's phase gauge field, the Chern number in the Brillouin zone corresponds to the winding number of $\theta(k_y)$, {\it i.e.}, $C_1=-\frac1{2\pi}\int_0^{2\pi}d\theta(k_y)$. For nontrivial $C_1$, the Wannier function $\ket{W(k_y,x)}$ is not periodic in $k_y$, since its center-of-mass position shifts by $C_1$ when $k_y$ is tuned continuously from $0$ to $2\pi$. In other words, the Wannier functions satisfy the following twisted boundary condition:
\begin{align}
\ket{W(k_y+2\pi,x)}=\ket{W(k_y,x+C_1)}\label{twistedBC}
\end{align}
As an example, we consider the lattice Dirac model\cite{qi2005,qi2008b} which is a two-band model with the Hamiltonian
\begin{align}
h({\bf k})&=\sum_{a=1,2,3}d_a({\bf k})\sigma^a\label{HDirac}
\end{align}
with $(d_1,d_2,d_3)({\bf k})=(\sin k_x,\sin k_y,M+2B(2-\cos k_x-\cos k_y))$. For $B>0$ and $-2<M/2B<0$, the system has Chern number $C_1=1$. The center-of-mass position of the Wannier functions can be obtained numerically as shown in Fig. \ref{fig1} (a), which shows the shift of Wannier function position under $k_y\rightarrow k_y+2\pi$. As a consequence of this shift, one can see intuitively from Fig. \ref{fig1} (a) that all Wannier functions $\ket{W(k_y,x)}$ can be parameterized by one real parameter. If we define $K_y=k_y+2\pi x$ for $k_y\in[0,2\pi)$, then
% for Chern number $C_1=1$ we see that each Wannier function $\ket{W(k_y,x)}$ can be obtained from an initial Wannier function $\ket{W(k_y=0,x=1)}$ by adiabatically tuning $k_y$. If we define
\begin{align}
\ket{W_{K_y=k_y+2\pi x}}\equiv \ket{W(k_y,x)}\label{Kydef}
\end{align}
is continuous in $K_y\in {\bf R}$. The center-of-mass position of $\ket{W_{K_y}}$ versus $K_y$ is shown in Fig. \ref{fig1} (b). %then all Wannier functions are labeled by one parameter $K_y\in {\bf R}$, and $\ket{W_{K_Y}}$ is continuous in $K_y$.
In this notion one can see clearly that the Wannier functions labeled by $K_y$ are analogous to the lowest Landau level wavefunctions of the ordinary QH state in the Landau gauge in the form of $\psi_{K_y}(x,y)={\left(\pi l_B^2L_y^2\right)^{-1/4}}e^{iK_yy-(x-K_yl_B^2)^2/2l_B^2}$,
which are also eigenstates of $k_y$ and localized in $x$ direction. Actually the Wannier functions $\ket{W_{K_y}}$ reduces exactly to the Landau level wavefunctions if we apply this formalism to the Hofstadter model\cite{hofstadter1976} and take the limit of small magnetic field $l_B\gg 1$.

Although the definition of $\ket{W_{K_y}}$ seems to be simply relabeling the Wannier functions, it plays a key role in understanding the QH and QAH states on equal footing. Once the analog of Landau level wavefunctions $\ket{W_{K_y}}$ is found, all many-body wavefunctions of FQH states can now find an analog in FQAH states.

{\it Laughlin states and pseudopotential Hamiltonians.--}

The FQAH analog of Laughlin states in FQH can be constructed by using the basis $\ket{W_{K_y}}$ to replace the Landau gauge wavefunctions in the Landau level problem. The Laughlin wavefunction on a cylinder is given by \cite{rezayi1994}
\begin{align}
\Psi_{1/m}({z_i})=\Omega\prod_{i<j}\left(e^{2\pi z_i/L_y}-e^{2\pi z_j/L_y}\right)^me^{-\sum_ix_i^2/2l_B^2}
\end{align}
with $\Omega$ a normalization factor, and $z_i=x_i+iy_i$. If we define the wavefunction in occupation number basis as
\begin{align}
\Phi(\left\{n_i\right\})=\frac{1}{L_y^N}\int\prod_i dx_idy_i\psi_{2\pi n_i/L_y}^*(x,y)\Psi_{1/m}({z_i})
\end{align}
then the FQAH version of the Laughlin state can be defined as
\begin{align}
\ket{1/m}&=\sum_{\left\{n_i\right\}}\Phi(\left\{n_i\right\})\prod_i\ket{W_{2\pi n_i/L_y}}
\end{align}
It is straighforward to verify that such a state is invariant under the lattice translation symmetry. In the occupation number basis, the wavefunction is the same as that of the Laughlin state, so that one can also define the FQAH version of the pseudo-potential Hamiltonian\cite{haldane1983,trugman1985}, for which the state $\ket{1/m}$ is a unique ground state. For example, for the $\nu=1/3$ Laughlin state the pseudo-potential Hamiltonian can be written in the following second-quantized form\cite{rezayi1994,lee2004}:
\begin{align}
H&=U\sum_{n\in{\bf Z}}b_n^\dagger b_n\label{H2d}\\
\text{with~}b_n&=\sum_{l\in{\bf Z},n-l\text{~even~}}\left(\frac l2e^{-\pi l^2/2L_y^2}\right)c_{\frac{n-l}2}c_{\frac{n+l}2}\nonumber
\end{align}
in which $c_n$ is the annihilation operator of the single particle state $\ket{W_{K_y=2\pi n/L_y}}$. In other words, $c_n^\dagger\ket{\phi}=\ket{W_{2\pi n/L_y}}$ with $\ket{\phi}$ the vacuum. The state $\ket{1/3}$ satisfies $b_n\ket{1/3}=0$, which is thus the ground state of $H$ (for $U>0$).

It is essential to show that the Hamiltonian (\ref{H2d}) is indeed a local Hamiltonian of the 2d lattice system. On this purpose it is convenient to consider the coherent states
\begin{align}
\ket{z}=\sum_{n\in{\bf Z}}e^{-i\frac{2\pi}{L_y}nz_2-\pi\left(z_1-\frac{n}{L_y}\right)^2}\ket{W_{{2\pi n}/{L_y}}}\label{coherentstate}
\end{align}
with $z=z_1+iz_2$ a complex variable and $z_1,z_2$ the real and imaginary parts. The coherent state $\ket{z}$ is periodic in $z\rightarrow z+iL_y$ which shows that the variable $z$ is defined on the cylinder. The coherent state is a superposition of $K_y$ eigenstates $\ket{W_{K_y}}$ around $K_y=2\pi n/L_y\simeq 2\pi z_1$, with the width of the distribution $\Delta K_y\simeq 2\sqrt{\pi}$. Since the $x$ position is proportional to $K_y$ as $x\simeq K_y/2\pi$, one can see that the coherent state is local (in the sense of exponential decay) in both $x$ and $y$ directions. For the two-band Hamiltonian (\ref{HDirac}), the coherent state wavefunction is shown in Fig. \ref{fig1} (d).

The annihilation operator of the coherent state can be defined as
\begin{align}
c(z)=\sum_{n\in{\bf Z}}e^{i\frac{2\pi}{L_y}nz_2-\pi\left(z_1-\frac{n}{L_y}\right)^2}c_n
\end{align}
Since the coherent state wavefunction is local, $c(z)$ is a local operator which is a superposition of the real space annihilation operators $c_i$ around the center-of-mass position $(z_1,z_2)$. The Hamiltonian (\ref{H2d}) can be written in $c(z)$ as
\begin{align}
H&=\frac{UL_y}{4\pi^2}\int dz_1dz_2{\bf b}^\dagger(z)\cdot {\bf b}(z),~{\bf b}(z)=c(z)(-i\nabla)c(z)\nonumber
\end{align}
Since $c(z)$ and ${\bf b}(z)$ are local operators in the 2d lattice model, so is $H$.

%Todo: fix the coefficients.

{\it Construction of more generic wavefunctions.--} The approach discussed above can be easily generalized to obtain more general FQAH states, such as the Moore-Read state\cite{moore1991} with non-Abelian quasiparticles. In general, any FQH state that can be written in the occupation number basis on the cylinder geometry has an analog in the FQAH system. If the FQH state is an exact ground state of some Hamiltonian, a similar Hamiltonian can be obtained for the FQAH state by using occupation number basis and/or coherent states. For example, in Ref. \cite{bernevig2008} a large class of FQH states have been constructed by using Jack polynomials. For such states the wavefunction in occupation number basis is recursively known, so that the generalization to FQAH states can be done straightforwardly.
%. Although the original construction in Ref. \cite{bernevig2008} is done on a plane geometry, it is straightforward to generalize the Jack polynomial states to cylinder geometry and then generalize it to the FQAH states. 
Similarly, other approaches of obtaining wavefunctions in FQH system ({\it e.g.} Ref. \cite{wen2008}) can also be generalized to FQAH system. Our result provides a systematical approach to obtain two-dimensional chiral topological states on a lattice model with full lattice translation symmetry and a small number of single particle states per unit cell. (On comparison, the ordinary FQH states can be considered as the FQAH states in a Hofstadter model in the continuum limit $l_B\gg a$ with $a$ the lattice constant and $l_B$ the magnetic length. In this limit the number of states per magnetic unit cell is diverging $\propto l_B^2/a^2$.) Our approach can also be further generalized to more generic QAH states with Chern number $C_1>1$, and/or multiple occupied bands.

%Add discussion on C1>1

%Up to now we have focused on the QAH states with Chern number $C_1=1$. One can also construct QAH states with $C_1>1$. For example, in the two-band model (\ref{HDirac}) if we define $(d_1,d_2,d_3)({\bf k})=(\cos k_x-\cos k_y,\sin k_x\sin k_y,M+2B(2-\cos k_x-\cos k_y))$,\cite{qi2005} the winding number of the mapping ${\bf k}\rightarrow {\bf d}({\bf k})$ is doubled, and we have $C_1=2$ for $-2<M/2B<0$. In this case the twisted boundary condition (\ref{twistedBC}) determines that the Wannier function shifts by $2$ in $x$ direction in the adiabatic evolution of $k_y$ from $0$ to $2\pi$. Consequently, the Wannier functions $\ket{W(k_y,x)}$ for even sites $x$ are connected by the boundary condition, but are disconnected from those with odd $x$. Similar to the definition of $\ket{W_{K_y}}$ in (\ref{Kydef}), for $C_1=2$ system we can group the Wannier functions into two groups:
%\begin{align}
%\ket{W_{K_y}^1}&=\ket{W(k_y,2x)},~\ket{W_{K_y}^2}=\ket{W(k_y,2x+1)}\nonumber\\
%\text{with~}K_y&=k_y+2\pi x,~x\in{\bf Z},~k_y\in[0,2\pi)
%\end{align}
%The wavefunctions of $\ket{W_{K_y}^1}$ and $\ket{W_{K_y}^2}$ are related by a lattice translation along $x$ direction. In this way, the QAH band with Chern number $C_1=2$ is represented by two states per wavevector $K_y$, which is similar to the single-particle wavefunctions in the Landau gauge for a bilayer QH system. Similar to the $C_1=1$ case, the wavefunctions of bilayer FQH system such as the Halperin states({\bf ref}) can be written in the occupation number basis and then generalized to the FQAH systems with $C_1=2$.

{\it Generalization to fractional topological insulators.--}

The approach of Wannier functions can also be further generalized beyond FQH states. In recent years, topological insulators (TI) have been proposed and experimentally realized, which can be considered as generalization of QH states in time-reversal invariant systems.\cite{TIReviews} In two-dimensions, TI can also be generalized to fractional TI\cite{bernevig2006a,levin2009}, which are fractionalized topological states protected by time-reversal symmetry. The simplest model for fractional TI consists of two decoupled FQH states, formed by spin up and down electrons, with opposite Hall conductance. For example for Laughlin state the corresponding fractional TI state is $\ket{FTI}=\ket{1/m,\uparrow}\otimes \ket{-1/m,\downarrow}$. More generally the spin $S_z$ conservation can be broken by spin-orbit coupling, and the TI and fractional TI states remain stable as long as time-reversal symmetry is preserved.\cite{kane2005b,levin2009} However, in the wavefunction approach it is difficult to construct a fractional TI wavefunction for a system without spin conservation, since the wavefunction of spin up (down) electrons in the spin-conserved case are holomorphic (anti-holomorphic) functions, and it is difficult to introduce a translation invariant coupling between them. The Wannier function approach provides a natural way to construct a fractional TI wavefunction for a generic Hamiltonian without spin conservation.

On this purpose we generalize the expression of maximally localized Wannier functions in Eq. (\ref{Wannier1d}) to a system with $N$ occupied bands\cite{yu2011}:
\begin{align}
\ket{W^i(k_y,x)}&=\frac1{\sqrt{L_x}}\sum_{k_x,m,n}e^{-ik_x\left({x-\frac{\theta_i(k_y)}{2\pi}}\right)}u^i_m
\nonumber\\
&\cdot\left[Pe^{-i\int_0^{k_x} a_x(p_x,k_y)dp_x}\right]_{nm}\ket{n,k_x,k_y}\label{Wannier1dmulti}
\end{align}
with $e^{i\theta_n(k_y)},~i=1,2,..N$ the eigenvalues of the Wilson loop operator $\left[Pe^{-i\int_0^{2\pi} a_x(p_x,k_y)dp_x}\right]$ and $u^i_m$ the corresponding eigenstates. $a_x$ is the $U(N)$ Berry's phase gauge field defined by $a_x(p_x,k_y)^{nm}=-i\bra{n,p_x,k_y}\frac{\partial}{\partial{p_x}}\ket{m,p_x,k_y}$. Similar to the one-band case, the center-of-mass position of the Wannier function $\ket{W^i(k_y,x)}$ is given by $\left\langle \hat{x}\right\rangle=x-\theta_i(k_y)/2\pi$. For a TI with a nontrivial $Z_2$ invariant\cite{kane2005b} in the occupied bands, the center-of-mass positions of the Wannier functions are doubly degenerate at time-reversal invariant wavevectors $k_y=0,\pi$, and there are two subgroups of Wannier functions with opposite odd winding number in the evolution of $k_y$ from $0$ to $2\pi$.\cite{yu2011,soluyanov2011b} As an example, we consider the Bernevig-Hughes-Zhang (BHZ) model for HgTe topological insulator\cite{bernevig2006c} and also include the bulk-inversion-asymmetry term\cite{koenig2008} such that the model is sufficiently generic and has no spin conservation. The lattice Hamiltonian can be written as
\begin{align}
h({\bf k})&=(M+2B(2-\cos k_x-\cos k_y))1\otimes \tau_z\nonumber\\
& +\sin k_x\sigma_z\otimes \tau_x+\sin k_y1\otimes\tau_y+\Delta \sigma_y\otimes\tau_y\label{BHZ}
\end{align}

\begin{figure}[htbp]
\centerline{\includegraphics[width=3.5in]{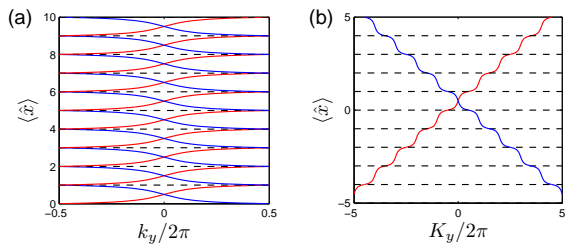}}
\caption{(Colour online) (a) The center-of-mass position $\left\langle\hat{x}\right\rangle$ of the Wannier functions versus $k_y$, and (b) $\left\langle\hat{x}\right\rangle$ versus the extended wavevector $K_y$ (see text), for the topological insulator model (\ref{BHZ}) with the parameters $M=-1,~B=1/2,~\Delta=0.2$. }\label{fig2}
\end{figure}
in which $\sigma_a,\tau_a$ are Pauli matrices in spin and orbital indices, and the time-reversal transformation is defined as $T=i\sigma_yK$ with $K$ the complex conjugation. The Wannier functions can be obtained for this model, as shown in Fig. \ref{fig2} (a). In the TI phase the two Wannier functions wind around the $k_y$ circle to opposite directions. Similar to Eq. (\ref{Kydef}), we can extend the definition of $k_y$ by continuity, and define the Wannier functions $\ket{W_{K_y}^{1,2}}=\ket{W^{1,2}(k_y,x_{1,2})}$, with $K_y=k_y\pm 2\pi x_{1,2}$ for $i=1,2$ respectively. The center-of-mass of these relabeled Wannier functions are shown in Fig. \ref{Kydef} (b). The important property of this Wannier function basis is that the occupied states are decoupled to two bands of Wannier states, each of which has the winding property as the Wannier states of a QAH system (with opposite winding direction), {\it even though the Hamiltonian does not preserve any spin conservation}. By using the Wannier functions $\ket{W^{1,2}_{K_y}}$ we can construct wavefunctions for fractional TI in the form of $\ket{FQAH_1}\otimes \ket{FQAH_2}$, which are direct product of FQAH state $\ket{FQAH_1}$ formed by $\ket{W^1_{K_y}}$ and the time-reversed state $\ket{FQAH_2}$ formed by $\ket{W^2_{K_y}}$. Although it appears to be a direct product state of two FQAH states, in the physical spin basis, spin up and down electrons are entangled and the state cannot be written as a direct product state. When a time-reversal invariant perturbation is considered (such as changing the term $\Delta$ in the Hamiltonian (\ref{BHZ})), the effect on the ansatz wavefunction can be described since the single particle Wannier states $\ket{W_{K_y}^{1,2}}$ depends on the perturbation. Thus we see that through the Wannier function approach one can obtain generic wavefunctions for fractional TI's without spin conservation.

I would like to thank Zheng-Cheng Gu and Shou-Cheng Zhang for helpful discussions. In particular I would like to thank Steven Kivelson for pointing out a mistake in the earlier version of the draft. This work is supported by the Alfred P. Sloan Foundation.

\bibliography{TI}

\end{document}